\documentclass{article}

\usepackage{arxiv}

\usepackage[utf8]{inputenc} 
\usepackage[T1]{fontenc}    
\usepackage{hyperref}       
\usepackage{url}            
\usepackage{booktabs}       
\usepackage{amsfonts}       
\usepackage{nicefrac}       
\usepackage{microtype}      
\usepackage{lipsum}
\usepackage{graphicx}
\graphicspath{ {./images/} }
\usepackage{natbib}
\usepackage{amsmath}
\usepackage{makecell}
\usepackage{adjustbox}
\usepackage[table]{xcolor}
\usepackage{enumitem}
\newtheorem{prop}{Proposition}

\def\bSig\mathbf{\Sigma}

\title{Power and Sample Size for \\ Marginal Structural Models}

\author{
 Bonnie E. Shook-Sa \\
 Department of Biostatistics \\ University of North Carolina at Chapel Hill \\ 
Chapel Hill, NC \\
  \texttt{bshooksa@live.unc.edu} \\
   \And
 Michael G. Hudgens \\
 Department of Biostatistics \\ University of North Carolina at Chapel Hill \\ 
Chapel Hill, NC \\
  \texttt{mhudgens@email.unc.edu} \\
}

\begin{document}
\maketitle
\begin{abstract}
Marginal structural models fit via inverse probability of treatment weighting are commonly used to control for confounding when estimating causal effects from observational data. When planning a study that will be analyzed with marginal structural modeling, determining the required sample size for a given level of statistical power is challenging because of the effect of weighting on the variance of the estimated causal means. This paper considers the utility of the \textit{design effect} to quantify the effect of weighting on the precision of causal estimates. The design effect is defined as the ratio of the variance of the causal mean estimator divided by the variance of a na\"{\i}ve estimator if, counter to fact, no confounding had been present and weights were not needed. A simple, closed-form approximation of the design effect is derived that is outcome invariant and can be estimated during the study design phase. Once the design effect is approximated for each treatment group, sample size calculations are conducted as for a randomized trial, but with variances inflated by the design effects to account for weighting. Simulations demonstrate the accuracy of the design effect approximation, and practical considerations are discussed.
\end{abstract}


\section{Introduction}
\label{s2:deffintro}

Researchers often aim to estimate causal effects rather than just associations between variables. In settings where experimental designs are implausible, inference relies on observational data from which measured associations can be confounded. Marginal structural models (MSMs) are a commonly used method to estimate causal effects in the presence of confounding variables \citep{hernan2000marginal,Robins2000,Cole2008,brumback2004sensitivity}. These models are fit using weighted estimating equations, where the weights are the inverse of each participant's probability of the observed treatment (or exposure). For a binary treatment, the estimand of interest is often the average causal effect, the difference in counterfactual means for the two treatment levels. With the assumptions of causal consistency, conditional exchangeability, and positivity, the inverse probability of treatment weight (IPTW) estimators are consistent for the MSM parameters for the causal means and the average causal effect \citep{Lunceford2004}. Variance estimates are computed using standard estimating equation theory \citep{Stefanski2002}, with the empirical sandwich variance estimator providing a consistent estimator for the asymptotic variance of the estimated average causal effect. 

While IPTW estimators provide researchers with an analytic tool for estimating causal effects in the presence of confounding variables, these estimators pose challenges during study design. The use of weights in the analysis affects the variance of the average causal effect estimator, making it challenging to determine the number of participants needed to achieve sufficient statistical power to detect a difference in causal means. Sample sizes cannot be calculated using standard methods that ignore weighting as in a randomized controlled trial (RCT) \citep[e.g. as in][]{chow2017sample}, as this will tend to be anti-conservative. Numerous papers have examined the properties of IPTWs and have developed guidelines and diagnostics for specifying weight models and adjusting estimated weights \citep{Austin2009,Austin2015,Cole2008,Lee2011}. However, currently no methods exist for power and sample size calculations for studies that will be analyzed using MSMs fit with IPTWs.

Weighted estimators are common in survey sampling and for Bayesian methods that utilize importance sampling, and both fields have developed methods to quantify the effect of weighting on the precision of estimates. \citet[page 257]{Kish1965} introduced the \textit{design effect} under the randomization-based inferential paradigm for survey sampling. The design effect is the ratio of the variance of an estimator under a complex sample design to the variance of the estimator under a simple random sample. When participants are selected directly from the finite population rather than from clusters of correlated observations, the design effect for a population mean estimator simplifies to the design effect due to weighting ($deff_w$), or the unequal weighting effect \citep{Kish1992}. Let $n$ be the sample size and $w_i$ represent the sampling weight for the $i^{th}$ participant, i.e., the inverse of participant $i$'s probability of selection. The design effect due to weighting is defined using either of the two equivalent forms: \begin{equation}
\label{eq:deffkish}
deff_w=\frac{n\sum_{i=1}^n{w_i}^2}{(\sum_{i=1}^n{w_i})^2} = 1+\frac{S^2(w)}{(n^{-1}\sum_{i=1}^n{w_i})^2}
\end{equation}
where $S^2(w)$ is the finite sample variance of the weights.
The design effect is interpreted as an estimator's increase in variance due to differential weights across participants. This metric is commonly applied to all types of complex sample designs in which individuals in the finite population have different probabilities of selection \citep[page 375]{Valliant2013}. \citet{Gabler} provided a justification for how Kish's design effect also applies to model-based estimators. In practice, the design effect is used to calculate the \textit{effective sample size}, which is equal to the observed sample size divided by the design effect. The effective sample size can be interpreted as the sample size under simple random sampling that that would have produced the same variance as the sample selected under the complex design \citep[page 5]{Valliant2013}. 

Bayesian importance sampling uses weighting methods when sampling from one distribution to estimate the properties of another distribution \citep{Kong1994}. Importance sampling uses the effective sample size metric to compare the precision of the weighted estimator to the precision that would be achieved if sampling had been conducted directly from the distribution of interest \citep{Kong1994}. When the estimator of interest is a H\'{a}jek estimator, \citet{Kong1992} provides an approximation for the effective sample size which is a function of (\ref{eq:deffkish}). 

Advantages of the approximated design effect are that it is outcome invariant and allows the sample size under a complex design to be translated into a sample size under a simpler design with the same variance. The former implies that the approximated design effect depends only on the participants' weights and is constant across outcomes. The latter means that once $deff_w$ is known or approximated, it can be used in power and sample size calculations along with the simpler assumptions needed to design a study without weights.  

In this paper we consider design effects for planning observational studies to assess the effect of a treatment or exposure on an outcome of interest. In the analysis of observational data, \citet{McCaffrey2004,McCaffrey2013} have used the effective sample size to quantify the loss of statistical precision following inference about causal effects using propensity score weighting. Here we describe the use of design effects for determining the sample size or power when designing an observational study. Section \ref{s2:deffdeff} introduces the design effect for causal inference and proves that it can be approximated with Kish's $deff_w$. Section \ref{s2:deffstudydes} demonstrates how the design effect can be used to determine the sample size or power of an observational study that will be analyzed using MSM with IPTWs. Section \ref{s2:deffsims} examines the accuracy of the design effect approximation for various exposure and outcome types via simulations, and Section \ref{s2:deffpractical} provides practical considerations regarding the use of design effects. Section \ref{s2:deffdiscuss} concludes with a discussion of the results and implications. The Appendix includes proofs of the propositions appearing in the main text. 

\section{The Design Effect}
\label{s2:deffdeff}

\subsection{Preliminaries}
Suppose an observational study is being planned where $n$ independent and identically distributed copies of $(A_i, L_i, Y_i)$ will be observed, where $A_i$ is the binary treatment (exposure) status for participant $i$  such that $A_i=1$ if participant $i$ received treatment and $A_i=0$ otherwise, $L_i$ is a vector of baseline covariates measured prior to $A_i$ or unaffected by treatment $A_i$, and $Y_i$ is the observed outcome for participant $i$. 

The aim of the observational study will be to estimate the effect of treatment $A$ on outcome $Y$. Specifically, let $Y_{1i}$ denote the potential outcome if an individual $i$, possibly counter to fact, receives treatment. Similarly let $Y_{0i}$ denote the potential outcome if individual $i$ does not receive treatment, such that $Y_i = A_i Y_{1i} + (1-A_i) Y_{0i}$. Inference from the observational study will focus on parameters of the MSM $E(Y_a) = \beta_0 + \beta_1 a$, with particular interest in the parameter $\beta_1$ which equals the average causal effect $ACE=E(Y_1)-E(Y_0)=\mu_1-\mu_0$. Note the MSM is saturated and thus does not impose any restrictions on the assumed structure of the data. 

Under certain assumptions, the parameters of the MSM can be consistently estimated using IPTW. In particular, assume conditional exchangeability holds, i.e., $Y_a \perp A \mid L$ for $a \in \{0,1\}$. Also assume that positivity holds such that $Pr(A=a \mid L=l)>0$ for all $l$ such that $dF_L(l)>0$ and $a \in \{0,1\}$, where $F_L$ is the cumulative distribution function of $L$. Estimating the average causal effect under the stated assumptions with the IPTW estimator first entails estimating the propensity score for each participant, defined as $p_i=Pr(A_i=1 \mid L_i)$ \citep{rosenbaum1983central}. A model is fit to obtain $\hat{p}_i$, each participant's estimated probability of treatment conditional on observed covariates $L_i$. The estimated IPTW is then equal to $\hat{W}_i=I(A_i=1)\hat{p}_i^{-1}+I(A_i=0)(1-\hat{p}_i)^{-1}$, where $I(A_i=a)$ is a \{0,1\} treatment indicator for participant $i$.  The estimated average causal effect $\hat{\beta_1}$ is obtained by regressing the observed outcome $Y$ on treatment $A$ with weights $\hat{W}$ using weighted least squares. The resulting IPTW estimator is a difference in H\'{a}jek estimators for the two causal means \citep{Hernan2010,Lunceford2004}: \begin{equation} \label{eq:diffhaj}
\widehat{ACE}=\hat{\mu}_1-\hat{\mu}_0=\frac{\sum_{i=1}^n{\hat{W}_i Y_i I\left(A_i=1\right)}}{\sum_{i=1}^n{\hat{W}_i I\left(A_i=1\right)}} - \frac{\sum_{i=1}^n{\hat{W}_i Y_i I\left(A_i=0\right)}}{\sum_{i=1}^n{\hat{W}_i I\left(A_i=0\right)}}\end{equation} 

Augmented IPW estimators, which incorporate both outcome and treatment models, may be used instead of (\ref{eq:diffhaj}) to estimate the $ACE$. Such estimators are doubly robust and will be more efficient than (\ref{eq:diffhaj}) if both the treatment and outcome models are correctly specified \citep{robins1994estimation, Lunceford2004}. Thus, the power and sample size calculations derived below, which are based on (\ref{eq:diffhaj}), will be conservative for studies analyzed with augmented IPW estimators. 

\subsection{The Design Effect for a Single Causal Mean}
Define the design effect to equal the ratio of the (finite sample) variance of $\hat{\mu}_a$ divided by the variance of a na\"{\i}ve causal mean estimator if, counter to fact, no confounding was present and weighting was not needed. That is, 
\begin{equation} 
\label{eq:deffdefinition}
{deff_w^a}=\frac{{Var}({\hat{\mu}_a)}}{{Var}({\tilde{\mu}_a)}}
\end{equation} where $\tilde{\mu}_a= \{ \sum_{i=1}^n Y_iI(A_i=a) \} / \{ \sum_{i=1}^n I(A_i=a) \}$. The derivation of the design effect estimator relies on the following proposition. The proposition assumes that the weights are known and are denoted by $W_a=P(A=a \mid L)^{-1}$ for $a \in \{0,1\}$ with $W=AW_1+(1-A)W_0$. Let $\sigma_a^2=Var(Y_a)$ for $a \in \{0,1\}$.
\begin{prop}
\label{Prop1}
\[\sqrt{n}(\hat{\mu}_a-\mu_a)\stackrel{d}{\rightarrow{}}N(0,\Sigma_a)\] where 
\[\Sigma_a=\sigma_a^2\left(\frac{E\left\{W^2I(A=a)\right\}}{{\left[E\left\{WI(A=a)\right\}\right]}^2}\right)+R(L, Y_a)
\] and \[R(L, Y_a)=E[\{W_a-E(W_a)\}(Y_a-\mu_a)^2]\] with \[|R(L, Y_a)| \leq \sqrt{Var(W_a)Var\{Y_a^2-2\mu_aY_a\}}\] for $a \in \{0,1\}$
\end{prop}
It follows from Proposition \ref{Prop1} that for large $n$ the variance of $\hat{\mu}_a$ can be approximated as:
\[Var(\hat{\mu}_a) \approx \frac{\sigma_a^2}{n}\left(\frac{E\left\{W^2I(A=a)\right\}}{{\left[E\left\{WI(A=a)\right\}\right]}^2}\right)+n^{-1}R(L, Y_a)
\] By similar arguments, for large $n$, $Var(\tilde \mu_a) \approx \sigma_a^2 / \{n P(A=a) \}$. Therefore,
\begin{equation}
\label{eq:eqdefffull}
deff_w^a \approx \frac{P(A=a) E\{W^2I(A=a)\}}{ {[E\{WI(A=a)\}]}^2}+Er_a\end{equation} where $Er_a=\{P(A=a) /\sigma^2_a \} R(L, Y^a)$, which by Proposition \ref{Prop1} is bounded by: \begin{equation} \label{eq:ERabound} |Er_a| \leq \{P(A=a) /\sigma^2_a \}\sqrt{Var(W_a)Var(Y_a^2-2\mu_aY_a)} \end{equation}

An approximation of (\ref{eq:eqdefffull}) that does not depend on the potential outcome $Y_a$ omits the remainder term $Er_a$: \begin{equation}
\label{eq:eqdefffullomitrem}
\widetilde{deff}_w^a = \frac{P(A=a) E\{W^2I(A=a)\}}{ {[E\{WI(A=a)\}]}^2}\end{equation} When planning an observational study, prior or pilot study data may be available to estimate (\ref{eq:eqdefffullomitrem}). In particular, suppose based on a pilot study $n_{p}$ copies of $(L_i,A_i)$ are observed. Then replacing $P(A=a)$ with $N_a / n_{p}$ where $N_a = \sum_{i=1}^{n_{p}} I(A_i=a)$, $E\{W^2I(A=a)\}$ with $n_{p}^{-1}\sum_{i=1}^{n_{p}}{\hat{W}_i}^2I(A_i=a)$,
and $E\{WI(A=a)\}$ with $n_{p}^{-1}\sum_{i=1}^{n_{p}}{\hat{W}_iI(A_i=a)}$, a consistent estimator of (\ref{eq:eqdefffullomitrem}) is: \begin{equation}
     \label{eq:deffestimator}
\widehat{deff}_w^a=\frac{N_a \sum_{i=1}^{n_p} \hat{W}_i^2 I(A_i=a)}{\big\{\sum_{i=1}^{n_p} \hat{W}_i I(A_i=a)\big\}^2} \end{equation}
\noindent{This estimator has the same form as Kish's design effect (\ref{eq:deffkish}), applied to treatment group $A=a$.} When prior data are not available, the design effect can be approximated using (\ref{eq:eqdefffullomitrem}) based on an assumed distribution for $A \mid L$ and the marginal distribution of $L$. The bias of (\ref{eq:eqdefffullomitrem}) or (\ref{eq:deffestimator}) as an approximation to (\ref{eq:eqdefffull}) in a given application depends on the value of $Er_a$. As further discussed in Section \ref{s2:deffdiscuss}, $Er_a$ is not guaranteed to be negligible. Bias of (\ref{eq:eqdefffullomitrem}) and (\ref{eq:deffestimator}) for varying outcome types and confounding structures is evaluated empirically in simulation studies presented in Section \ref{s2:deffsims}.

\section{Sample Size Calculations using the Design Effects}
\label{s2:deffstudydes}

When the $ACE$ is the focus of inference for the observational study being planned, the large sample distribution of $\widehat{ACE}$ can be used for power or sample size calculations. As $n \to \infty$, $\widehat{ACE}$ is consistent and asymptotically normal, i.e., $\sqrt{n}(\widehat{ACE}-ACE)\stackrel{d}{\rightarrow{}}N(0,\Sigma^*)$, where $\Sigma^*$ is given by equation (13) in \citet{Lunceford2004}. By the following proposition, $\Sigma^*$ can be decomposed into the sum of asymptotic variances for the two causal mean estimators:

\begin{prop}
\label{PropACE}
\[\Sigma^*=\Sigma_1+\Sigma_0\]
\end{prop}

Treating the weights as fixed or known leads to a larger asymptotic variance for $\widehat{ACE}$ compared to appropriately treating the weights as estimated, i.e., $\Sigma^*$ is at least as large as the true asymptotic variance of $\widehat{ACE}$ \citep{Lunceford2004}. Therefore, sample size formulae derived based on $\Sigma^*$ would  in general be expected to be conservative.

The results in Propositions \ref{Prop1} and \ref{PropACE} allow for sample size calculations for studies that will be analyzed using MSM with IPTW. Suppose the sample size for the observational study being planned is to be determined on the basis of the power to test $H_0: ACE=0$ versus $H_a: ACE \neq 0$ or equivalently $H_0: \beta_1=0$ versus $H_a: \beta_1 \neq 0$. Define the test statistic $t = \widehat{ACE} \{Var(\widehat{ACE})\}^{-1/2}$, where \begin{equation} \label{eq:varACE}
Var(\widehat{ACE}) \approx Var(\hat{\mu}_1) + Var(\hat{\mu}_0) \approx \{n P(A=1)\}^{-1} \sigma_{1,adj}^2+\{n P(A=0)\}^{-1} \sigma_{0,adj}^2
\end{equation} with $\sigma^2_{a,adj}=\sigma^2_a deff_w^a$ for $a \in \{0, 1\}$. Then, 
\begin{equation}
\label{eq:zstat} \frac{\widehat{ACE}-ACE}{\sqrt{Var(\widehat{ACE})}}
\end{equation} is approximately standard normal for large $n$. Thus, $H_0$ is rejected when $|t|>z_{1-\frac{\alpha}{2}}$, where $\alpha$ is the type I error rate and $z_q$ is the $q^{th}$ quantile of the standard normal distribution.

\begin{prop}
\label{Prop2}
The sample size required to achieve power $1-\beta$ for effect size $ACE=\delta$ and type I error rate $\alpha$ is approximately:
\begin{equation}
\label{eq:samplesize}
n_{deff} = \frac{(1+k)(z_{1-\alpha / 2}+z_{1-\beta})^2(\sigma_{1,adj}^2 / k + \sigma_{0,adj}^2)}{\delta^2}
\end{equation}
where $k=P(A=1) / P(A=0)$ is the odds of treatment in the population. 
\end{prop}

The sample size formula (\ref{eq:samplesize}) is the standard sample size equation commonly used to design RCTs, but with $\sigma_a^2$ replaced by $\sigma_{a,adj}^2$ \citep[][page 48]{chow2017sample}. Thus, Proposition \ref{Prop2} simplifies power and sample size calculations for observational studies by allowing researchers to design studies as if they were designing an RCT, but inflating the assumed variances by the approximated design effects. The researcher first assumes that no confounding is present, specifies the desired $\alpha$ and $1-\beta$, and makes assumptions about $\sigma_0^2$, $\sigma_1^2$, $\delta$, and $k$. The design effect is then approximated. When data from a pilot or prior study are available, $deff_w^1$ and $deff_w^0$ can be approximated based on (\ref{eq:deffestimator}) for each treatment group. When no prior study data are available, the distribution of the anticipated weights can be estimated based on assumptions about the distribution of $L$ and $A \mid L$ and the design effect can be calculated based on (\ref{eq:eqdefffullomitrem}). While these assumptions may not be easy to make, this approach notably requires no assumptions about the potential outcomes $Y_0$ and $Y_1$ and their associations with $A$ and $L$. Further discussion about these practical considerations is included in Section \ref{s2:deffpractical}. Once the design effects are approximated by $\widetilde{deff}_w^a$ or $\widehat{deff}_w^a$, adjusted variances $\sigma_{a,adj}^2$ can be estimated by $\tilde{\sigma}_{a,adj}^2=\sigma_a^2\widetilde{deff}_w^a$ or $\hat{\sigma}_{a,adj}^2=\sigma_a^2\widehat{deff}_w^a$, respectively, for $a \in \{0,1\}$.

\section{Simulation Study}
\label{s2:deffsims}

\subsection{Simulation Scenarios}
Simulation studies were conducted to demonstrate use of the design effect in study design and estimate the bias of the approximation in (\ref{eq:eqdefffullomitrem}) and (\ref{eq:deffestimator}) under a variety of confounding structures and outcome types. The scenarios in Table \ref{tab:Table1} were considered. For all scenarios, $\alpha=0.05$ and $1-\beta=80\%$ were chosen.

\begin{table}[ht]
\caption{Five simulation scenarios. Scenarios 1-4 demonstrate use of the design effect when no prior study data are available, and Scenario 5 demonstrates use of the design effect with prior study data. $X \sim B(p)$ indicates that a random variable $X$ follows the Bernoulli distribution with probability of success equal to $p$.}
\centering
\normalsize
\setlength{\tabcolsep}{4pt} 
\renewcommand{\arraystretch}{1.5} 
\begin{tabular}{c c c c c c} 
\hline

  & Scenario & Exposure ($A$) & Confounders ($L$) & Outcome ($Y$) & $\delta$ \\ \hline 
  \\
  1& \makecell{binary $Y$, \\ small $deff_w^a$} & \makecell{$A \mid L=0 \sim B(0.5)$ \\ $A \mid L=1 \sim B(0.75)$} & $L \sim B(0.6)$ & \makecell{$Y_0 \mid L \sim B(0.85-0.2L)$ \\ $Y_1 \mid L \sim B(0.70-0.2L)$} & $-0.15$ \\
  \\

 2& \makecell{binary $Y$, \\ large $deff_w^a$} & \makecell{$A \mid L=0 \sim B(0.1)$ \\ $A \mid L=1 \sim B(0.9)$} & $L \sim B(0.5)$ & \makecell{$Y_0 \mid L \sim B(0.85-0.2L)$ \\ $Y_1 \mid L \sim B(0.70-0.2L)$} & $-0.15$ \\
 \\
 3& \makecell{continuous $Y$, \\ small $deff_w^a$} & \makecell{$A \mid L=0 \sim B(0.5)$ \\ $A \mid L=1 \sim B(0.75)$} & $L \sim B(0.6)$ & \makecell{$Y_0 \mid L \sim N(20-10L, 144)$ \\ $Y_1 \mid L \sim N(25-10L, 256)$} & $5.0$ \\
 \\

 4& \makecell{continuous $Y$, \\ large $deff_w^a$} & \makecell{$A \mid L=0 \sim B(0.1)$ \\ $A \mid L=1 \sim B(0.9)$} & $L \sim B(0.5)$ & \makecell{$Y_0 \mid L \sim N(20-10L, 144)$ \\ $Y_1 \mid L \sim N(25-10L, 256)$} & $5.0$ \\
 \\
 5& \makecell{prior study data \\ (NHEFS)} & \makecell{smoking cessation} & \makecell{9 baseline \\ variables} & weight gain & $2.0$ \\
\hline
\end{tabular}
\label{tab:Table1}
\end{table}

\subsection{Sample Size Calculation}
Two general approaches can be used to design a study with the design effect approximation: when prior study data are not available, as in Scenarios 1-4, and when prior study data are available, as in Scenario 5. One example from each general approach is presented in detail.

\subsubsection{Example 1: No prior study data (Scenario 1)} 

Suppose no prior study data are available to design the study of interest. Then, the researcher must make the same assumptions and design choices as when designing an RCT, namely by specifying $\alpha$, $1-\beta$, $\sigma_0^2$, $\sigma_1^2$, $\delta$, and $k$. In general, $\sigma_1^2$ can be determined by deriving the marginal distribution of $Y_1$ based on the assumed distributions of $Y_1 \mid L$ and $L$. For Scenario 1, 
$P(Y_1=1) = \sum_{l=0}^1 P(Y_1=1 \mid L=l) P(L=l) = 0.58$, and thus $\sigma_1^2=0.2436$. Similarly, $\sigma_0^2=0.1971$. Here, the average causal effect is assumed to be $\delta=-0.15$. The proportion of the population receiving treatment can be derived by integrating the distribution of $A \mid L$ over $L$. For Scenario 1, 
$P(A=1)= \sum_{l=0}^1 P(A=1 \mid L=l) P(L=l) = 0.65$, and thus $k \approx 1.857$. 
When prior study data are not available, the distribution of the IPTWs must be assumed at the design phase. Based on the assumptions in Table \ref{tab:Table1}, four possible values of $W$ exist. These assumed values of $W$, along with the joint distribution of $A$ and $L$, allow for the computation of the design effects using (\ref{eq:eqdefffullomitrem}). This leads to $\widetilde{deff}_w^0 = 1.12$ and $\widetilde{deff}_w^1 = 1.04$, with approximated adjusted variances of $\tilde{\sigma}_{0,adj}^2=0.2208$ and $\tilde{\sigma}_{1,adj}^2=0.2533$.

Under the assumptions outlined in Table \ref{tab:Table1} for Scenario 1, to achieve 80\% power to detect an average causal effect of $-0.15$ at the $\alpha=0.05$ level, a sample size of approximately $n_{deff}=356$ is required based on Proposition \ref{Prop2}. The design effects and required sample sizes for Scenarios 2-4 can be determined similarly and are presented in Table \ref{tab:Table3}. Note Scenarios 1 and 3 have the same design effects because in both instances the joint distribution of $A$ and $L$ is the same. Likewise, Scenarios 2 and 4 have the same design effects.

\subsubsection{Example 2: Prior study data (Scenario 5)} 

Prior study or pilot data may allow for better informed assumptions about $\sigma_0^2$, $\sigma_1^2$, $\delta$, and $k$. Because $\sigma_a^2=E(Y_a^2)-\{E(Y_a)\}^2$, $\sigma_a^2$ can be estimated by obtaining $\hat{E}(Y_a^2)$ and $\hat{E}(Y_a)$ from fitted MSMs based on the prior study data. The estimate $\widehat{ACE}$ and prevalence of the exposure or treatment in the prior study can inform assumptions about $\delta$ and $k$. 

As an example, consider designing a new study based on the National Health and Nutrition Examination Survey Data I Epidemiologic Follow-up Study (NHEFS) example presented in Chapter 12 of \citet{Hernan2010}. Hern{\'a}n and Robins use MSM with IPTWs to estimate the average causal effect of smoking cessation ($A$) on weight gain after approximately 10 years of follow-up ($Y$) based on the NHEFS sample of smokers ($n=1566$), assuming conditional exchangeability based on nine baseline confounders $L$: sex, age, race, education, smoking intensity, duration of smoking, physical activity, exercise, and weight. 

Making the same assumptions as \cite{Hernan2010}, Scenario 5 considers the design of a new study to estimate the average causal effect of smoking cessation on 10-year weight gain. Based on the NHEFS data, assume that $\sigma_0^2=56.1$ and $\sigma_1^2=74.0$, obtained by fitting MSMs with IPTWs to estimate $E(Y_a^2)$ and $E(Y_a)$. In the Hern{\'a}n and Robins example, $\widehat{ACE}=3.441 kg$. The new study will be designed to provide approximately 80\% power to detect a difference in weight gain of $\delta=2.0kg$. From the NHEFS sample, assume $k \approx 0.346$.

When prior study data are available, $deff_w^0$ and $deff_w^1$ can be estimated using (\ref{eq:deffestimator}). For the NHEFS data, $\widehat{deff}_w^0=1.03$ and $\widehat{deff}_w^1=1.24$. This leads to approximated adjusted variances of $\hat{\sigma}_{0,adj}^2=57.78$ and $\hat{\sigma}_{1,adj}^2=91.76$. Based on these assumptions, a sample size of $n_{deff}=853$ is needed to achieve approximately 80\% power to detect an average causal effect of $2.0 kg$ at the $\alpha=0.05$ level using MSM with IPTWs.

\subsubsection{Na\"{\i}ve Sample Size Calculations} 
As a comparison, sample sizes $n_{rct}$ were calculated naively under the assumptions of an RCT, ignoring the effect of weighting on the variances of the estimates. In other words, sample sizes were calculated as demonstrated above, except using $\sigma_a^2$ instead of $\tilde{\sigma}_{a,adj}^2$ or $\hat{\sigma}_{a,adj}^2$ from Table \ref{tab:Table3}. 

\begin{center}
\begin{table}[ht]
\centering
\caption{Variances, approximated design effects, approximated adjusted variances, and required sample sizes for simulation scenarios by treatment.}
\setlength{\tabcolsep}{10pt} 
\renewcommand{\arraystretch}{2.5} 
\begin{tabular}{c c c c c c c c} 
\hline

  & Scenario & a & $\sigma_a^2$ & $\widetilde{deff}_w^a$ or $\widehat{deff}_w^a$ & $\tilde{\sigma}_{a,adj}^2$ or $\hat{\sigma}_{a,adj}^2$ & $n_{deff}$ & $n_{rct}$ \\
 \hline
 1& \makecell{binary $Y$, \\ small $deff_w^a$} & \makecell{$0$ \\ $1$} & \makecell{$0.1971$ \\$0.2436$} &  \makecell{$1.12$ \\ $1.04$} & \makecell{$0.2208$ \\ $0.2533$} & $356$ & $327$\\

 2& \makecell{binary $Y$, \\ large $deff_w^a$} & \makecell{$0$ \\ $1$} & \makecell{$0.1875$ \\ $0.2400$} &  \makecell{$2.78$ \\ $2.78$ } & \makecell{$0.5208$ \\ $0.6667$ } & $828$ & $298$\\

 3& \makecell{continuous $Y$, \\ small $deff_w^a$} & \makecell{$0$ \\ $1$ } & \makecell{$168.0$ \\ $280.0$} &  \makecell{$1.12$ \\ $1.04$} & \makecell{$188.2$ \\ $291.2$} & $310$ & $286$\\

 4& \makecell{continuous $Y$, \\ large $deff_w^a$} & \makecell{$0$ \\ $1$} & \makecell{$169.0$ \\ $281.0$} &  \makecell{$2.78$ \\ $2.78$} & \makecell{$469.4$ \\ $780.6$} & $784$ & $283$ \\

 5& \makecell{prior study data, \\ (NHEFS)} & \makecell{$0$ \\ $1$} & \makecell{$56.10$ \\ $74.00$} &  \makecell{$1.03$ \\ $1.24$} & \makecell{$57.78$ \\ $91.76$} & $853$ & $713$\\
\hline
\end{tabular}
\label{tab:Table3}
\end{table}
\end{center}

\subsection{Evaluation}
For Scenarios 1-4, empirical power based on samples of size $n_{deff}$ was evaluated via simulation by following these steps:

\begin{enumerate}[label=(\roman*)]
	\item Generate a superpopulation of size $N=1,000,000$ based on distributions in Table \ref{tab:Table1}.
	\item \label{step2} Select a sample of size $n_{deff}$ without replacement from the superpopulation, where $n_{deff}$ is specified in Table \ref{tab:Table3}.
	\item \label{step3} Estimate $\hat{W}_i$ for each member of the sample based on the predicted values from the logistic regression of $A$ on $L$.
	\item Fit the MSM $E(Y_{ai})=\beta_0+\beta_1 a_i$ using weighted least squares, treating the weights as estimated by stacking the estimating equations from the weight model with the estimating equations for the causal means and difference in causal means using the geex package in R \citep{saul}.
	\item \label{step5} Test $H_0: \beta_1=0$ versus $H_1: \beta_1 \ne 0$ using a Wald test, rejecting $H_0$ at the $\alpha=0.05$ significance level. 
	\item Repeat steps \ref{step2}-\ref{step5} $R=2000$ times and calculate empirical power as the proportion of simulated samples where $H_0$ was rejected.
\end{enumerate}

For Scenario 5, empirical power based on a sample of size $n_{deff}$ was evaluated via simulation by following these steps:
\begin{enumerate}[label=(\roman*)]
	\item Estimate the propensity score for each of the $1566$ NHEFS participants from a logistic regression model of $A$ on $L$ as $\hat{p}_i=\widehat{Pr}(A_i=1 \mid L_i=l_i)$. As in \citet{Hernan2010}, the logistic regression model includes main effects for each of the nine baseline confounders and quadratic terms for the four continuous covariates.
	\item For each participant, calculate $\hat{Y}_{ai}$, $a \in \{0,1\}$, as the predicted value $\hat{E}(Y_{ai} \mid L_i=l_i)$ from the following linear regression model, fit only on participants with $A=a$: $E(Y_{ai} \mid L_i=l_i)=l_i\beta$, where $l_i$ is a vector for participant $i$ that includes an intercept term, the 9 previously defined covariates, and the four quadratic terms corresponding to continuous covariates. Also compute $\widehat{Var}(Y_{ai} \mid L_i=l_i)=MSE_a$, where $MSE_a$ is the mean squared error from the model for $E(Y_{ai})$.
	\item Add $1.441$ to $\hat{Y}_{0i}$ for all participants, such that $ACE=2.0$ in the simulated population instead of $3.441$ as in the NHEFS sample.
	\item \label{step2Scen5} Select a sample of size $n_{deff}$ with replacement from the NHEFS dataset, where $n_{deff}$ is specified in Table \ref{tab:Table3}.
	\item Assign $A_i=a_i$ as a random draw from $A_i \sim Bernoulli(\hat{p}_i)$.
	\item Let $Y_{ai}=\hat{Y}_{ai}+\epsilon_{ai}$, where $\epsilon_{ai} \sim N(0,\widehat{Var}(Y_{ai} \mid L_i=l_i))$.
	\item \label{step5Scen5} Follow steps \ref{step3}-\ref{step5} from the above list for Scenarios 1-4.
	\item Repeat steps \ref{step2Scen5}-\ref{step5Scen5} $R=2000$ times and estimate empirical power as the proportion of simulated samples where $H_0$ was rejected.
\end{enumerate} For each scenario, these steps were repeated to calculate empirical power based on the na\"{\i}ve sample sizes, replacing $n_{deff}$ with $n_{rct}$.

The results of the simulation study are presented in Table \ref{tab:Table4}. For all simulation scenarios, when the sample size was calculated using the design effect, empirical power was equal to or exceeded the nominal 80\% level. That is, use of the design effects to calculate required sample sizes led to close to the intended level of statistical power. On the other hand, ignoring the effect of weighting and basing sample sizes on the na\"{\i}ve assumptions of an RCT led to empirical power that was lower than the nominal 80\% level for all but one scenario. These results demonstrate that ignoring the weights in power and sample size calculations can lead to significantly underpowered studies, particularly when there are strong confounders that lead to high variability in the weights.

For all scenarios, the approximation errors $Er_a$ from (\ref{eq:eqdefffull}) for each sample and treatment were estimated by $\widehat{Er}_a=\{N_a / (n\sigma_a^2) \} 
\hat{E}\left[\{\hat{W_a}-\hat{E}(\hat{W_a})\}\{Y_a-\hat{E}(Y_a)\}^2\right]$, where expected values were calculated empirically within each sample. Estimated approximation errors were then averaged across the $R=2000$ simulated samples. Mean estimated approximation error was small for most scenarios (Table \ref{tab:Table4}) and was in opposite directions for the two treatment groups, which tended to offset the effects of the errors. Approximation error was large for Scenario 2 ($0.60$ for $A=0$ and $-0.19$ for $A=1$), but empirical power still equaled the nominal level when the design effect was used to calculate the sample size. Note Scenario 2 is an extreme example, as it includes only a single and very strong confounding variable and only two possible and extreme values for $W$. For the binary outcome, this resulted in large approximation errors.

\begin{center}
\begin{table}[ht]
\centering
\caption{Results of the simulation study by scenario across $R=2000$ samples. Empirical power $n_{deff}$ and $n_{rct}$ are the proportions of simulated samples in which the p-values for testing $H_0: \beta_1=0$ versus $H_1: \beta_1 \ne 0$ were less than $\alpha=0.05$ for the following MSM: $E(Y_{ai})=\beta_0+\beta_1 a_i$, based on sample sizes $n_{deff}$ and $n_{rct}$, respectively, from Table \ref{tab:Table3}} 
\setlength{\tabcolsep}{10pt} 
\renewcommand{\arraystretch}{3} 
\begin{tabular}{c c c c c} 
\hline

  Scenario & \makecell{Empirical \\ Power \\ $n_{deff}$} & \makecell{Empirical \\ Power \\ $n_{rct}$} & \makecell{Mean \\ $\widehat{Er}_0$} & \makecell{Mean \\ $\widehat{Er}_1$} \\
 \hline
1	&	0.81	&	0.76	&	0.08	&	-0.01	\\
2	&	0.80	&	0.42	&	0.60	&	-0.19	\\
3	&	0.85	&	0.81	&	-0.02	&	0.01	\\
4	&	0.86	&	0.47	&	0.00	&	-0.01	\\
5	&	0.82	&	0.76	&	0.02	&	-0.03	\\

\hline
\end{tabular}

\label{tab:Table4}
\end{table}
\end{center}

\section{Practical Considerations}
\label{s2:deffpractical}
When prior study data are not available, specifying the design effects can be challenging. A few general guidelines are offered to help researchers determine reasonable assumptions to facilitate power and sample size calculations. 

When only a few categorical covariates will be included in the weight model, researchers can use subject matter knowledge or prior study information to nonparametrically specify the joint distribution of $A$ and $L$, or the marginal distribution of $L$ and the conditional distribution of $A \mid L$ (as in Example 1). Based on these assumptions, the anticipated weights can be calculated nonparametrically and the design effects for each treatment group can be approximated.

When specification of these distributions is not feasible, researchers can forgo approximating the values of the weights and instead consider more generally how much variation is expected in the weights. The lower bound for $deff_w^a$ is 1, which implies that the weights within both treatment groups are all equal and thus covariates are not predictive of the treatment. Design effects tend to increase when more covariates are added to the weight model. The presence of covariates that are strong predictors of treatment tends to increase the design effect. Care must be taken to identify the appropriate set of confounders to include in the weight model \citep{Vansteelandt2012}. Inclusion of instrumental variables, which are predictive of the exposure but which do not affect the outcome, inflate the variance of the $ACE$ estimator without reducing bias \citep{rubin1997estimating,myers2011effects}. The use of weight truncation will decrease the design effect. 

\begin{figure}
\begin{center}
  \includegraphics[width=\linewidth]{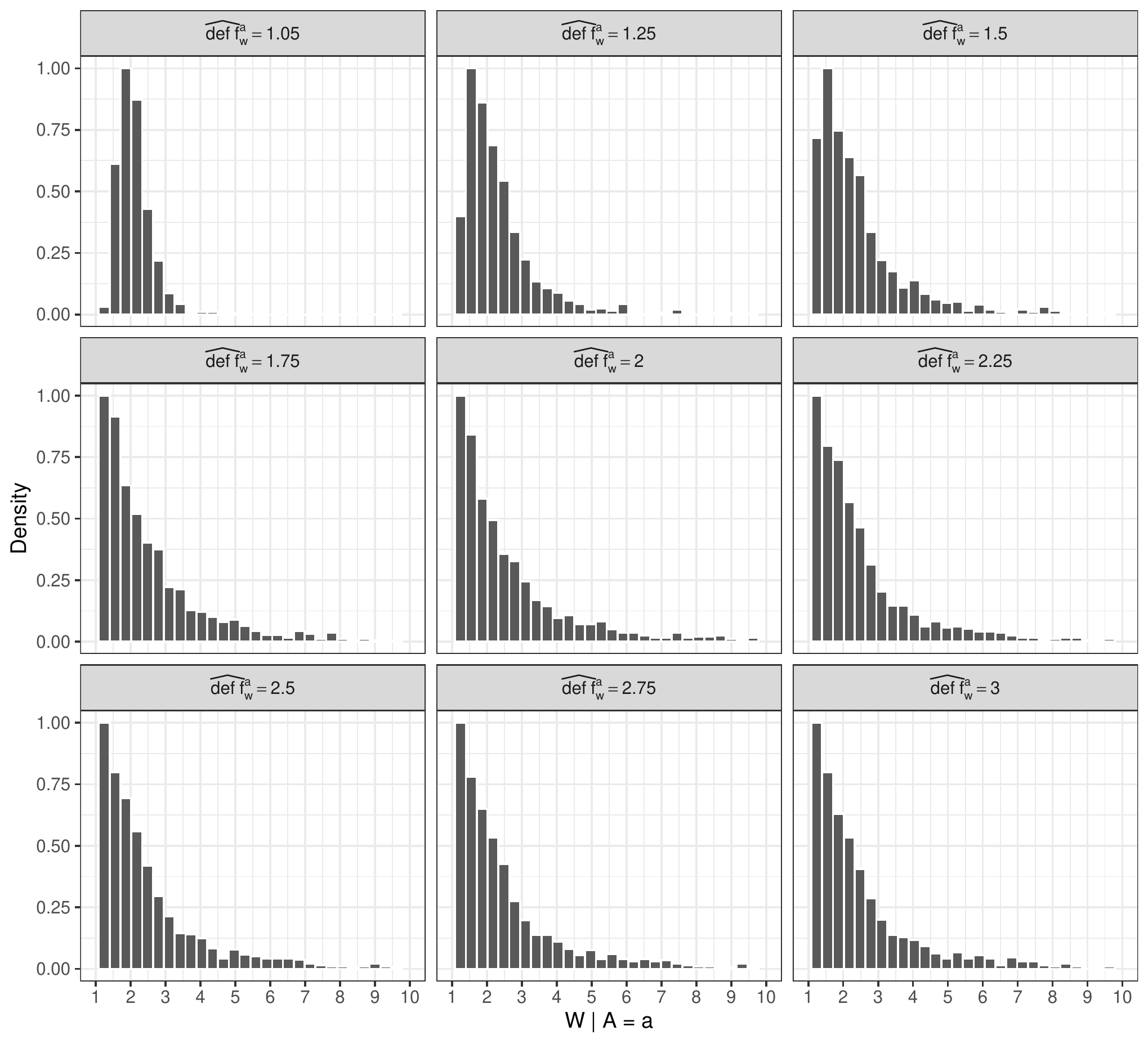}
  \caption{Examples of weight distributions for various approximated design effects. Distributions were generated by taking the reciprocals of $N_a=1000$ random draws from beta distributions with mean 0.5 and shape parameters set to achieve the desired design effect.}
  \label{fig:deffscenarios}
\end{center}
\end{figure}

Figure \ref{fig:deffscenarios} provides a visual depiction of weight distributions within one treatment group for various values of the design effect to aid researchers in choosing a design effect consistent with the expected variation in the weights. These weight distributions were generated by taking the reciprocals of $N_a=1000$ random draws from beta distributions with mean 0.5 and shape parameters set to achieve the desired design effect. As variation in the weights increases, so does the design effect approximation. 

\section{Discussion}
\label{s2:deffdiscuss}

The design effect approximation simplifies power and sample size calculations of observational studies. Using the design effect allows researchers to utilize standard power and sample size software (e.g., nQuery, SAS Proc Power) for randomized trials, but with variances inflated by the approximate design effects. An additional advantage of using the design effect approximation is that no assumptions are required about the relationship between the potential outcomes and either the treatment or the confounders. Empirical results presented in Section \ref{s2:deffsims} demonstrate the design effect approximation can yield the nominal level of power over a range of confounding and outcome structures.  

Approximating the design effect when planning an observational study may be challenging. In survey sampling, it is common practice to report estimated design effects in analytic reports for better understanding of the precision of the estimates and to assist other researchers who are designing similar studies \citep[see, for example][]{NSDUH}. Reporting the estimated design effects corresponding to treatment or exposure effect estimates in observational studies may assist researchers with future study designs. In time, as more studies analyzed with MSMs start to report their design effects, rules of thumb and practical upper bounds for the design effects will likely emerge to aid in the design of future studies (see, for example, \citet[page 41]{united2008designing}, \citet[page 251]{daniel2011sampling}, and \cite{salganik2006variance} from the survey sampling literature).

In the absence of knowledge of estimated design effects from prior studies, the design effect may be approximated either using (\ref{eq:eqdefffullomitrem}) or, if pilot data are available, (\ref{eq:deffestimator}). In either case, the remainder term in (\ref{eq:eqdefffull}) is ignored, which may introduce bias. The remainder may be large when individuals with extreme weight values tend to have potential outcomes that are also extreme relative to the mean. In the simulation studies in Section \ref{s2:deffsims}, the approximation error was small for all but one of the scenarios examined. Remainders were in opposite directions for the two treatment groups, which tended to offset the effects of the errors and thus use of the approximation did not result in deviations from the nominal level of statistical power for any of the scenarios examined. However, there is no guarantee that approximation error will be negligible for a given study. When pilot or prior study data are available, approximation error $Er_a$ can be estimated as in the simulations, but with $Y_a$ replaced with $\hat{Y}_a$ for $a \in \{0,1\}$ where $\hat{Y}_a$ is based on an assumed outcome regression model. Alternatively, an estimate for the upper bound of $Er_a$ can be obtained by estimating the upper bound in (\ref{eq:ERabound}).

Despite these limitations, the design effect approximation can be a useful tool for the design of studies that will be analyzed using MSM with IPTWs, as currently no power and sample size methods exist in this context. The design effect can also be used in precision calculations using approaches analogous to those described in this paper, i.e., basing calculations on the adjusted variances $\tilde{\sigma}^2_{a,adj}$ or $\hat{\sigma}^2_{a,adj}$ rather than $\sigma^2_{a}$.

\section*{Acknowledgements}
The authors thank Stephen Cole, Noah Greifer, Bryan Blette, Kayla Kilpatrick, Shaina Alexandria, and Jaffer Zaidi for their helpful suggestions. This work was supported by NIH grant R01 AI085073. The content is solely the responsibility of the authors and does not necessarily represent the official views of the NIH.\vspace*{-8pt}

\bibliographystyle{plainnat}
\bibliography{references}  

\appendix
\setcounter{equation}{0}
\renewcommand{\theequation}{A.\arabic{equation}}
\section*{Appendix}
\label{s:app}
\subsection*{Proof of Proposition \ref{Prop1}}
Without loss of generality, consider $a=1$. Let $X_i=W_{1i} Y_i A_i$ and $Z_i=W_{1i} A_i$. The asymptotic distribution of $\hat{\mu}_1=\sum_{i=1}^n X_i/\sum_{i=1}^n Z_i$ can be derived using the multivariate delta method \citep{Kong1992}. Let \[T_n=\left(\begin{array}{cc}
\frac{1}{n}\sum_{i=1}^{n_{\ }}X_i \\
\frac{1}{n}\sum_{i=1}^{n_{\ }}Z_i
\end{array}\right),\ \ \ \theta=\left(\begin{array}{cc}
\mu_x \\
\mu_z
\end{array}\right), \ \ \  g(\theta)=\frac{\mu_x}{\mu_z},\ \ \ \nabla{}g(\theta)=\left(\begin{array}{cc}\frac{1}{\mu_z} \\
-\frac{\mu_x}{\mu_z^2} \end{array}\right),\]
and \[\Sigma{}=\left(\begin{array}{cc}
Var(X) & Cov(X,Z) \\
Cov(Z,X) & Var(Z)
\end{array}\right)\] where $\mu_x=E\left(X_i\right)$, $\mu_z=E\left(Z_i\right)$, and $\nabla{}g(\theta)$ is the gradient vector for $g(\theta)$. From the bivariate central limit theorem,
$\sqrt{n}(T_n-\theta)\stackrel{d}{\rightarrow{}}N_2(0,\Sigma{})$. Applying the multivariate delta method, $\sqrt{n}\{g(T_n)-g(\theta)\}\stackrel{d}{\rightarrow}N(0,\Sigma_1)$ where $g(T_n)=\hat{\mu}_1$ and \[\Sigma_1=\nabla{}g(\theta)^{T}\Sigma{}\nabla{}g(\theta) = 
{\left(\frac{\mu{}_x}{\mu{}_z}\right)}^2\left\{\frac{Var\left(X\right)}{\mu{}_x^2}+\frac{Var(Z)}{\mu{}_z^2}-2\frac{Cov\left(X,Z\right)}{\mu{}_x \mu{}_z}\right\}
\] \noindent{Dropping} subscripts $i$ for notational ease, note that: \begin{equation*} \label{eq:EWA}
\mu{}_z=E\left(Z\right)=E\left(W_1A\right)=E_{L}\left\{\frac{E_{A \mid L}A}{P(A=1 \mid L)}\right\}=1
\end{equation*} and from \cite{Hernan2010} Technical Point 2.3, $\mu{}_x=E\left(X\right)=E\left(W_1AY_1\right)=\mu_1$. Then, \begin{equation} 
\label{eq:sigma1init}
\Sigma_1=Var\left(W_1AY_1\right)+\mu_1^2Var\left(W_1A\right)-2{\mu_1}Cov\left(W_1AY_1,W_1A\right)
\end{equation} A simpler form for $\Sigma_1$ is derived by rewriting the components of (\ref{eq:sigma1init}) using the following results. First note that \begin{align}
Cov\left(W_1AY_1,W_1A\right)&=E\left\{\left(W_1AY_1\right)\left(W_1A\right)\right\}-E\left(W_1AY_1\right)E(W_1A) \nonumber 
=E\left(W_1^2AY_1\right)-\mu_1 \nonumber \\ &= E_{L}\left\{\frac{E_{A \mid L}A{\ \ 
E}_{Y_1 \mid L}Y_1}{{P(A=1 \mid L)}^2}\right\}-\mu_1 \nonumber =E_{L}\left\{\frac{{\ \ \ E}_{Y_1 \mid L}Y_1}{{P(A=1 \mid L)}^{\
}}\right\}-\mu_1 \nonumber \\
\label{eq:Cov(WAY,WA)}
&=E\left(W_1Y_1\right)-\mu_1 
\end{align}

Also note that \begin{align}
Var\left(W_1AY_1\right)&=E\left(W_1^2AY_1^2 \right)-{\left\{E\left(W_1AY_1\right)\right\}}^2 \nonumber \\
&=E_{L}\left\{\frac{E_{A \mid L}A{\ \ \
E}_{Y_1 \mid L}Y_1^2}{{P(A=1 \mid L)}^2}\right\}-\mu_1^2 \nonumber =E_{L}\left\{\frac{{\ \ \ E}_{Y_1 \mid L}{Y_1^2}}{{P(A=1 \mid L)}}\right\}-\mu_1^2  \nonumber
\\ \label{eq:VarWAY}
&=E\left(W_1 Y_1^2\right)-\mu_1^2
\end{align}
By the law of total variance:
\begin{equation}
\label{eq:VarWA}
Var(W_1A)=E\{Var(W_1A \mid L)\}+Var\{E(W_1A \mid L)\} =E(W_1)-1
\end{equation} Therefore, plugging (\ref{eq:Cov(WAY,WA)}), (\ref{eq:VarWAY}), and (\ref{eq:VarWA}) into (\ref{eq:sigma1init}),
\begin{align}
\Sigma_1&=E(W_1Y_1^2)-\mu_1^2+\mu_1^2\{E(W_1)-1 \}-2\mu_1\{E\left(W_1Y_1\right)-\mu_1\} \nonumber \\ \label{eq:prelim1}
&=E(W_1Y_1^2)+\mu_1^2E(W_1)-2\mu_1E(W_1Y_1)  
\end{align}
Next define $R=E[\{W_1-E(W_1)\}(Y_1-\mu_1)^2]$ and note that
\begin{equation} \label{eq:Remdef}
R=E(W_1Y_1^2)-2\mu_1E(W_1Y_1)-E(Y_1^2)E(W_1)+2\mu_1^2E(W_1)
\end{equation}
From (\ref{eq:prelim1}) and (\ref{eq:Remdef}) it follows that
\begin{align}
\Sigma_1&=E(W_1)E(Y_1^2)-E(W_1)\mu_1^2+R \nonumber 
= E(W_1)\sigma_1^2 +R \nonumber \\
&= \sigma_1^2E(W_1^2A)+R \nonumber 
=\sigma_1^2\left[\frac{E(W_1^2A)}{E(W_1A)}\right]+R \nonumber \\
&=\sigma_1^2\left[\frac{E(W^2A)}{\{E(WA)\}^2}\right]+R \nonumber
\end{align}

Bounds for $R$ follow from the Cauchy-Schwarz inequality: \[|R|=|Cov(W_1,Y_1^2-2\mu_1Y_1)| \leq \sqrt{Var(W_1)Var(Y_1^2-2\mu_1Y_1)}\] 

\subsection*{Proof of Proposition \ref{PropACE}}
From equation (13) in \citet{Lunceford2004}, \[\Sigma^*=E \{W_1(Y_1-\mu_1)^2+ W_0(Y_0-\mu_0)^2\}\]
Note
\begin{align*} \label{eq:LDsigma} E \{ W_1(Y_1-\mu_1)^2 \} &=  E(W_1Y_1^2)-2\mu_1E(W_1Y_1)+\mu_1^2E(W_1) \end{align*}
which equals $\Sigma_1$ by (\ref{eq:prelim1}). Similarly, $E \{ W_0(Y_0-\mu_0)^2 \}=\Sigma_0$, proving the proposition.

\subsection*{Proof of Proposition \ref{Prop2}}
Let $1-\beta$ denote the power to detect a difference in causal means of size $\delta$, i.e., 
\begin{align*}
1-\beta &= P(|t|>z_{1-\alpha/2} \mid ACE=\delta) \\
&= P \left(\frac{\widehat{ACE}-\delta}{\sqrt{Var(\widehat{ACE})}} > z_{1-\alpha/2} - \frac{\delta}{\sqrt{Var(\widehat{ACE})}} \biggr\rvert ACE=\delta \right) \\
& \quad + P \left(\frac{\widehat{ACE}-\delta}{\sqrt{Var(\widehat{ACE})}} < z_{\alpha/2} - \frac{\delta}{\sqrt{Var(\widehat{ACE})}} \biggr\rvert ACE=\delta \right)
\end{align*}

In large samples, (\ref{eq:zstat}) is approximately standard normal. Thus,
\begin{equation} \label{eq:power}
1-\beta \approx 1-\Phi \left(z_{1-\alpha/2} - \frac{\delta}{\sqrt{Var(\widehat{ACE})}} \right) + \Phi \left( z_{\alpha/2} - \frac{\delta}{\sqrt{Var(\widehat{ACE})}}  \right )
\end{equation} where $\Phi(*)$ represents the cumulative distribution function for the standard normal evaluated at $*$. Without loss of generality, assume $\delta>0$. Then the second component on the right side of (\ref{eq:power}) will be less than $\alpha/2$ and often close to zero. Therefore, 
\begin{equation} \label{eq:derivesmpsize} z_{\beta} \approx z_{1-\alpha/2} - \frac{\delta}{\sqrt{Var(\widehat{ACE})}} \end{equation} Define $k=P(A=1)/P(A=0)$. Given that $Var(\widehat{ACE}) \approx \{n P(A=1)\}^{-1} \sigma_{1,adj}^2+\{n P(A=0)\}^{-1} \sigma_{0,adj}^2$ and solving (\ref{eq:derivesmpsize}) for $n$ yields (\ref{eq:samplesize}).

\end{document}